\documentclass[journal]{IEEEtran}

\usepackage{ifpdf}
\usepackage{graphicx}
\DeclareGraphicsExtensions{.pdf,.jpg,.png}
\usepackage[cmex10]{amsmath}
\usepackage{amssymb}
\usepackage{array}
\usepackage{mdwmath}
\usepackage{mdwtab}
\usepackage{pslatex}
\usepackage{xcolor}
\usepackage{textcomp}
\usepackage{placeins}
\usepackage{eqparbox}
\usepackage{url}
\usepackage{stfloats}
\usepackage{multicol}
\usepackage{float}
\usepackage{algorithmicx}
\usepackage{algpseudocode}
\usepackage{amsthm}
\usepackage{comment}
\usepackage{cite}
\usepackage{color, soul}
\usepackage{enumerate}
\usepackage{bbm}
\usepackage{threeparttable}
\usepackage{booktabs}
\usepackage{array}
\usepackage{balance}
\usepackage{color}
\usepackage{gensymb}
\usepackage[font=footnotesize]{subcaption}
\usepackage[labelsep=period]{caption}

\begin{document}

\title{Hover or Perch: Comparing Capacity of Airborne and Landed Millimeter-Wave UAV Cells}

\author{Vitaly Petrov$^1$, Margarita Gapeyenko$^1$, Dmitri Moltchanov, Sergey Andreev, and Robert W. Heath Jr.
\thanks{V. Petrov, M. Gapeyenko, D. Moltchanov, and S. Andreev are with Tampere University, Finland. R. Heath is with The University of Texas at Austin, USA.}
\thanks{$^1$A part of this work was completed during the research visit of Vitaly Petrov and Margarita Gapeyenko to The University of Texas at Austin, USA. This work was supported in part by the Academy of Finland (project RADIANT) and by the National Science Foundation under Grant No. ECCS-1711702.}}

\maketitle
\begin{abstract}
On-demand deployments of millimeter-wave (mmWave) access points (APs) carried by unmanned aerial vehicles (UAVs) are considered today as a potential solution to enhance the performance of 5G+ networks. \textcolor{black}{The battery lifetime of modern UAVs, though, limits the flight times in such systems.} In this letter, we evaluate a feasible deployment alternative for temporary capacity boost in the areas with highly fluctuating \textcolor{black}{user demands. The approach} is to land UAV-based mmWave APs on the nearby buildings instead of hovering over the area. Within the developed mathematical framework, we compare the system-level performance of \emph{airborne} and \emph{landed} deployments by taking into account the full operation cycle of the employed drones. Our numerical results demonstrate that the choice of the UAV deployment option is determined by an interplay of the separation distance between the service area and the UAV charging station, drone battery lifetime, and the number of aerial APs in use. The presented methodology and results can support efficient on-demand deployments of UAV-based mmWave APs in prospective 5G+ networks.\end{abstract}
\begin{IEEEkeywords}
mmWave networks, drone cells, beyond-5G.
\end{IEEEkeywords}

\IEEEpeerreviewmaketitle

\section{Introduction}
\label{sec:intro}

Millimeter-wave (mmWave) communication is one of the key features introduced by the fifth-generation (5G) wireless networks. The use of mmWave bands enables transmissions with higher data rates than those available in 4G microwave systems. With the first cellular networks exploiting mmWave connectivity \textcolor{black}{already appearing}, follow-up research is targeting advanced topologies, such as multi-hop mmWave communication and moving mmWave cells~\cite{mmWave_survey}. Among these opportunities, the use of mobile mmWave access points (APs) mounted on unmanned aerial vehicles (UAVs) is an attractive solution to temporarily boost network capacity and coverage~\cite{mmWave_uav_magazine2}.

These UAV-based mmWave APs enable on-demand network densification \textcolor{black}{to facilitate e.g., spontaneous massive events} and thus provide additional flexibility on top of the static infrastructure. Without a major innovation in battery technology, UAVs will continue to suffer from battery constraints~\cite{related_uav_coverage}. In addition, drones hovering over people's heads raise safety concerns and noise pollution from the UAV motors.

\textcolor{black}{An} alternative deployment option for the UAV-based mmWave APs is to land the drones on the objects surrounding the service area rather than continuously fly over it. Modern drones can land on rooftops, balconies, lampposts, and even perch to building walls without the operator involvement~\cite{perch_uav_1}. Following this method, the UAV-based mmWave APs can be temporarily deployed in the service area and leave their landing point only to recharge batteries or when the service is no longer needed. \textcolor{black}{Such levels of flexibility exceed those that could be provided by the static network infrastructure alone.}

\textcolor{black}{As compared to the \emph{airborne} UAV-based APs, these} \emph{landed} deployment {options result in a considerable improvement to the battery lifetime, since the primary source of energy consumption -- \textcolor{black}{the motors} -- can be switched off during the AP service time. At the same time, the mmWave access link becomes a challenge as the appropriate landing point may be farther away from the target user equipment (UE) than the AP location under \emph{airborne} option, where the drone can hover closer to the UEs. The greater separation distances consequently result in lower signal-to-noise (SNR) levels~\cite{margo_tvt_mc}.

The \emph{landed} and \emph{airborne} deployment options have their own strengths and weaknesses. Depending on the system parameters, either option can be preferable. Therefore, a thorough comparison of airborne vs. landed alternatives for UAV-based mmWave APs in prospective usage scenarios is needed.

\textcolor{black}{Some studies on landed and airborne drones have been done in prior work.} The authors in~\cite{related_heath_conference} analyzed the concept where drones deploy microwave APs in the service area. Later, the work in~\cite{related_spatial_framework} contributed a spatial framework for airborne mmWave networks. Finally, methods to account for UAV battery \textcolor{black}{constraints} in airborne microwave networks were presented in~\cite{related_uav_coverage} and~\cite{revision_energy_all}, among others. To the best of our knowledge, no comparison of airborne vs. landed deployment options for UAV-based mmWave APs has been offered to date. We aim to address this gap in the present work.

{
\color{black}
In this letter, we compare the network performance provided by \emph{airborne} and \emph{landed} deployments of UAV-based mmWave APs. We first develop a mathematical framework that models a full UAV operating cycle -- deployment, service, charging, and travel to/from the service area -- as well as accounts for mmWave-specific propagation. We then apply our framework to compare the performance of the considered options.
}

\section{System Model}
\label{sec:system_model}

The model focuses on \textcolor{black}{serving a spontaneous massive event in} an unobstructed circular service area (e.g., a city square) of radius $R$ and features a drone charging station at a distance of $\ell$ from the area (see Fig.~\ref{fig:system_model}). The service area is populated with humans distributed randomly by following the Poisson point process (PPP) with density $\lambda$. Each human carries a mmWave UE at height $h_{\text{U}}$ \textcolor{black}{and distance $r_{\text{U}}$ from the body. Human bodies are modeled as cylinders of height $h_{\text{B}}$, $h_{\text{B}}>h_{\text{U}}$, and radius $r_{\text{B}}$.}

\subsection{Modeling mmWave Communications}\label{sec:propagation}
Each of the mmWave UEs is connected to the nearest UAV-based mmWave AP. We assume a fixed AP Tx power, $P_{\text{A}}$, and directional antennas at the AP and UE sides with the gains of $G_{\text{A}}$ and $G_{\text{U}}$. \textcolor{black}{To facilitate a first-order analysis, we assume that directional mmWave communications in our setup are primarily noise-limited.} \textcolor{black}{As the service area represents an open square, the mmWave link between the UE and its nearest UAV-based AP can be occluded by human bodies, but not by buildings. Hence, we model the line-of-sight (LoS) mmWave link, as either \emph{blocked LoS} (due to self-blockage or blockage caused by other pedestrians) or \emph{non-blocked LoS}~\cite{margo_tvt_mc,3gpp_los}.}

The received power on a \textcolor{black}{\emph{non-blocked LoS}} link to the UE from the nearest UAV-based AP, $P_{\text{N}}$, is modeled by
\begin{align}\label{eq:pathloss}
P_{\text{N}} = P_{\text{A}} G_{\text{A}} G_{\text{U}} A_{\text{N}} d^{-\gamma},
\end{align}
where $d$ is the 3D distance between the nodes \textcolor{black}{and $A_{\text{N}}$ is the attenuation coefficient for non-blocked LoS propagation.} \textcolor{black}{Since the altitude of UAV-based mmWave APs is assumed to not exceed several tens of meters, we follow the parameters from the 3GPP UMi-Street Canyon model and set $\gamma = 2.1$, $A_{\text{N}} = 10^{-3.24}f^{-2}_{C}$, where $f_{\text{C}}$ is the carrier frequency~\cite{3gpp_los}.}
 
If the LoS path is occluded by humans \textcolor{black}{(\emph{blocked LoS} link)}, the received power degrades according to the measurements in~\cite{haneda_blockage_vtc}: $P_{B} = P_{\text{A}} G_{\text{A}} G_{\text{U}} A_{\text{B}} d^{-\gamma}$, where \textcolor{black}{$A_{\text{B}}$ is the attenuation coefficient for blocked LoS propagation ($A_{\text{B}} \ll A_{\text{N}}$).} 

\subsection{Modeling UAV Operation Cycle}\label{sec:drone_cycle}
We analyze and compare two options for the UAV operation: \emph{airborne} and \emph{landed}\footnote{\textcolor{black}{To highlight the trade-offs between the UAV deployment options in a clear setup, we do not model stationary mmWave APs. The framework can be further extended to consider the presence of ground network infrastructure.}}. In the airborne case, drones hover above the \textcolor{black}{service area. In the landed case,} drones are assumed to land on or perch to the nearby buildings~\cite{perch_uav_1}. For both options, the fleet comprises of $N$ UAV-based mmWave APs. Each UAV follows its operation cycle with three major stages: (i) \emph{serving stage}, when the drone provides wireless access to the UEs; (ii)~\emph{charging stage}, when the drone charges its battery at the charging station; and (iii) \emph{en-route stage}, when the drone is flying from the charging station to the service area or back.

Following the described cycle (see Fig.~\ref{fig:system_model}), only a certain subset of UAV-based APs \textcolor{black}{are} available for serving users: $N_{\text{A}} < N$ and $N_{\text{L}} < N$ for airborne and landed options, respectively. At the serving stage, all the UAVs are deployed at the same heights: $h_{\text{A}}$ for the airborne option and $h_{\text{L}}$ for the landed option. In both cases, the drone travels to the charging station when its remaining energy is just enough for this flight. 

\subsection{UAV Locations During Service}
Recalling that the SNR is a decreasing function of distance to the UE, the task of deploying $N_{\text{A}}$ airborne APs is reduced to minimizing the sum of distances to all the potential UEs~\cite{davies1979cluster}. Hence, for the circular shape of the service area, we use circle in circle packing as a simple yet rational approximation~\cite{circle_packing_saad}. Exact deployments following the circle in circle packing exist for each practical number of UAV-based APs ($N_{\text{A}}<7$).

Following similar considerations, $N_{\text{L}}$ landed drones are assumed to be located at the circumference of the circle having the radius $R$ at the equal distances from each other, \textcolor{black}{thus approximating the UAV deployment on some of the buildings surrounding the service area (see Fig.~\ref{fig:system_model})}.

\begin{figure}[t!]
	\centering
	\includegraphics[width=0.89\columnwidth]{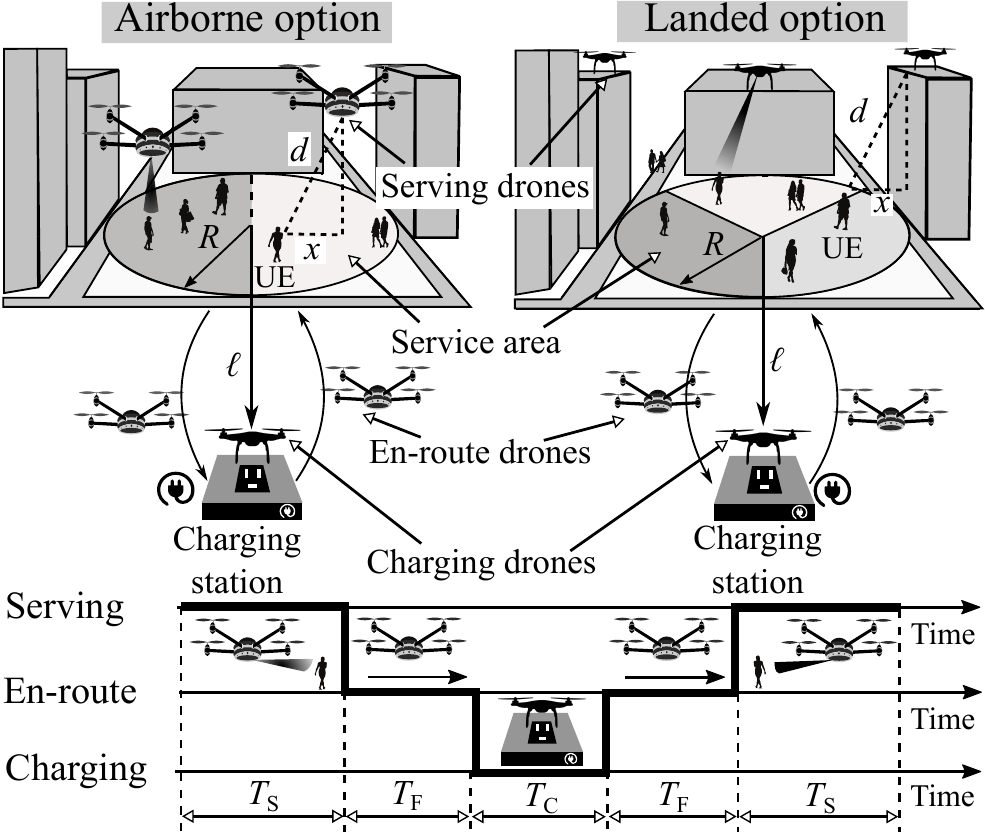}
	\caption{\textcolor{black}{Illustration of the considered system model.}}
	\label{fig:system_model}
\end{figure}

\section{Evaluation Methodology}
\label{sec:methodology}


\subsection{Projections of mmWave Links}\label{sec:links_projections}
{
\color{black}
Let us first assume that there are currently $M$ UAV-based mmWave APs within the service area. Each of these $M$ APs thus serves the UEs in a sector of radius $R$ and angle $2\pi/M$. 

\subsubsection{General approach}
\label{subsubsec:general}
To obtain the probability density function (PDF) of distances between a random UE and its nearest AP for both \emph{landed} and \emph{airborne} options, we utilize the conventional stochastic geometry approach~\cite{revision_link_projection}. Particularly, the PDF of the 2D distance $x$ is produced as the length of the arc of radius $x$ around the AP projection onto $\Re^2$ lying within the considered sector divided by the area of this sector.

\subsubsection{Airborne option}
Here, we apply the approach detailed above for the airborne option. We offer an example for $M=5$ below. PDFs for other values of $M$ are obtained similarly.

For $M=5$, the radius of the packed circle is $r_{\text{A}}=R/(1+\sqrt{2+2/\sqrt{5}})$~\cite{davies1979cluster}\textcolor{black}{, thus determining the 2D distance between the AP ($A$ in Fig.~\ref{fig:airborne}) and the nearest border of the sector.}


Following the approach in subsection~\ref{subsubsec:general} and the geometry in Fig.~\ref{fig:airborne}, the PDF of the 2D distance between a random UE and its nearest UAV-based AP, $X_{\text{A}}$, has three branches
\begin{align}\label{eqn:flyingProjection}
f_{X_{\text{A}}}(x)=
\begin{cases}
2\pi{}x/Q,&\hspace{-2mm}0 \leq x < r_{\text{A}},\\
[2\pi{}x-L_1(x)-2L_2(x)]/Q,&\hspace{-2mm}r_{\text{A}} \leq x < d_{\text{A}},\\
L_3(x)/Q,&\hspace{-2mm}d_{\text{A}} \leq x \leq R-r_{\text{A}},\\
\end{cases}
\end{align}
where $L_1(x)$, $L_2(x)$, and $L_3(x)$ are the arc lengths in Fig.~\ref{fig:airborne}, $Q=\pi{}R^{2}/M$, \textcolor{black}{while $d_{\text{A}}$ is the 2D distance between the AP and the corner of the sector shown in Fig.~\ref{fig:airborne3} and calculated in (\ref{eq:d_a})}.

Notice that $\alpha=\pi/5$, $\psi=(\pi-\alpha)/2$. Applying the cosine theorem to $d_{\text{A}}=||AF||$ (see Fig.~\ref{fig:airborne}), we obtain
\begin{align}
\label{eq:d_a}
d_{\text{A}}=\sqrt{r_{\text{A}}^{2}+\left(2R\sin\alpha/2\right)^{2}-4r_{\text{A}}R\sin\alpha/2\cos\psi}.
\end{align}
Consequently, from basic geometry in Fig.~\ref{fig:airborne}, we have
\begin{align}
&L_{1}(x)=2x\left[2\pi-\cos^{-1}\left(\frac{x^{2}+(R-r_{\text{A}})^{2}-R^2}{2x(R-r_{\text{A}})}\right)\right],\\
&L_{2}(x)=2x\cos^{-1}\left( \left[x^{2}+z^{2}_{\text{A}}-(R-r_{\text{A}})^{2}\right] / 2xz_{\text{A}} \right),
\end{align}
\begin{align}
&z_{\text{A}}=(R-r_{\text{A}})\cos\alpha-\sqrt{(R-r_{\text{A}})^{2}(\cos^{2}\alpha-1)+x^{2}},\\
&L_{3}(x)=2x\left(\pi-\cos^{-1}\left[\frac{x^{2}+z^{2}_{\text{A}}-(R-r_{\text{A}})^{2}}{2xz_{\text{A}}}\right]-\frac{\pi}{5}\right).
\end{align}

\begin{figure}[t!]
\centering
\begin{subfigure}[b]{0.25\columnwidth}
		\centering
		\includegraphics[height=3.3cm]{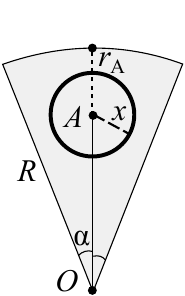}
	\caption{\textcolor{black}{The first branch}}
	\label{fig:airborne1}
	\end{subfigure}
\hfil
	\begin{subfigure}[b]{0.41\columnwidth}
		\centering
		\includegraphics[height=3.3cm]{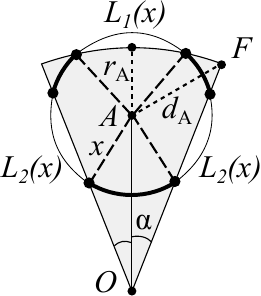}
	\caption{\textcolor{black}{The second branch}}
	\label{fig:airborne2}
	\end{subfigure}
\hfill
	\begin{subfigure}[b]{0.3\columnwidth}
		\centering
		\includegraphics[height=3.3cm]{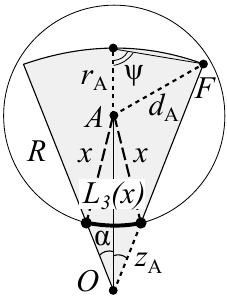}
	\caption{\textcolor{black}{The third branch}}
	\label{fig:airborne3}
	\end{subfigure}
\caption{\textcolor{black}{Link length pdf branches for airborne option, $f_{X_{\text{A}}}(x)$}.}
\label{fig:airborne}
\end{figure}

\subsubsection{Landed option}
Recall that the \emph{landed} UAVs are located at the circumference of the serving area. The PDF of the 2D distance between a random UE and its nearest UAV-based AP, $X_{\text{L}}$, is obtained by following the approach detailed above.

Particularly, for $M \geq 3$, we have
\begin{align}\label{eqn:coord1}
f_{X_{\text{L}}}(x) =
\begin{cases}
[2x\cos^{-1}(x/2R)]/Q, &0 \leq x < r_{\text{L}}, \\
[2x\cos^{-1}(x/2R) - L_{4}(x)]/Q, &r_{\text{L}} \leq x < d_{\text{L}}, \\
L_{5}(x)/Q, &d_{\text{L}} \leq x \leq R, \\
\end{cases}
\end{align}
where $\beta=\pi/M$ and the parameters are given by
\begin{align}
\begin{split}
&L_{4}(x) = 4x \cos^{-1}([R\sin\beta]/x),\quad{}d_{\text{L}} = R\sqrt{2 - 2\cos\beta},\\
&L_{5}(x)\hspace{-0.5mm}=\hspace{-0.5mm}x\left[\pi - 2\beta  - 2\cos^{-1}\left(R\sin\beta/x\right)\right],\quad{}r_{\text{L}}\hspace{-0.5mm}=\hspace{-0.5mm}R\sin\beta.
\end{split}
\end{align}
The PDFs of $X_{\text{L}}$  for $M = 1$ and $M = 2$ are obtained similarly.
}

\subsection{UAV Flight and Charging Processes}

\subsubsection{Landed option} The fraction of time when the tagged drone is in the \emph{serving} stage, $\rho_{\text{L}}$, can be approximated as $T_{\text{S},\text{L}}/(T_{\text{S},\text{L}}+2T_{\text{F}}+T_{\text{C}})$, where $T_{\text{S},\text{L}}$ is the time in service, $T_{\text{C}}$ is the charging time, and $T_{\text{F}}$ is the time required to fly from the charging station to the service area.

{
\color{black}
Let us define $P_{\text{E}}$ as the power of the UAV motors in watts, $\nu$ as the drone cruise speed during the \emph{en-route} stage\footnote{\textcolor{black}{The exact value of $P_{\text{E}}$ depends not only on the drone flight speed, $\nu$, but also on multiple other parameters: drone design, size, weight, etc. Later, in the numerical study, we follow the approximation for $P_{\text{E}}$ given in~\cite{revision_energy_all}.}}, $E$ as the UAV battery capacity in watt-hours, and $P_{\text{T}}$ as the UAV-based AP power in watts\footnote{\textcolor{black}{$P_{\text{E}} \gg P_{\text{T}} \gg P_{\text{A}}$. $P_{\text{T}}$ includes the power of all the transceiver chains, signal processing, and data processing at the UAV-based mmWave AP.}}. The energy budget is thus}
\begin{align}\label{eq:energy_budget_s}
T_{\text{S},\text{L}}P_{\text{T}} + 2T_{\text{F}}P_{\text{E}} = E.
\end{align}
Writing $T_{\text{F}} = \ell / \nu$ and substituting $T_{\text{S},\text{L}}$ from (\ref{eq:energy_budget_s}), we obtain
\begin{align}
\rho_{\text{L}} = \frac{T P_{\text{E}} \nu - 2P_{\text{E}}\ell}{T P_{\text{E}} \nu + 2\ell(P_{\text{T}}-P_{\text{E}}) + T_{\text{C}} \nu P_{\text{T}}},
\label{eq:rho_s}
\end{align}
where $T$ is the drone flight time on batteries and $E = T P_{\text{E}}$.

\subsubsection{Airborne option}
\textcolor{black}{The main difference here is that the UAV consumes $P_{\text{T}} + P_{\text{H}}$ instead of $P_{\text{T}}$ when in service, where $P_{\text{H}}$ is the engine power consumption in the hovering regime\footnote{\textcolor{black}{In our numerical study, we calculate $P_{\text{H}}$ following the approach from~\cite{revision_energy_all}.}}. Hence, the drone energy budget can be expressed as}
{
\color{black}
\begin{align}\label{eq:energy_budget_f}
T_{\text{S},\text{A}}(P_{\text{T}}+P_{\text{H}}) + 2T_{\text{F}}P_{\text{E}} = E,
\end{align}
}
where $T_{\text{S},\text{A}}$ is the duration of the serving stage.

Consequently, the fraction of time when the drone remains in the serving stage for the airborne option, $\rho_{\text{A}}$, is derived as
{
\color{black}
\begin{align}
\rho_{\text{A}} = \frac{T P_{\text{E}} \nu - 2P_{\text{E}}\ell}{T P_{\text{E}} \nu + 2 \ell \left(   P_{\text{T}} + P_{\text{H}} - P_{\text{E}}    \right) + T_{\text{C}} \nu (P_{\text{H}}+P_{\text{T}})}.
\label{eq:rho_f}
\end{align}
}

\subsubsection{Number of drones serving the area}
We now use the values $\rho_{\text{L}}$ and $\rho_{\text{A}}$ to estimate the number of drones that are simultaneously in service. Following~\cite{parker1996deterministic}, we set $N_{\text{L}} = \lfloor N \rho_{\text{L}} \rfloor$ and $N_{\text{A}} = \lfloor N \rho_{\text{A}} \rfloor$, so that $N_{\text{L}}$ and $N_{\text{A}}$ represent the number of serving drones \emph{guaranteed} to be present in the service area. We utilize $N_{\text{A}}$ and $N_{\text{L}}$ below to determine the metrics of interest.

\subsection{Capacity Analysis}
\subsubsection{Spectral efficiency}

We first obtain the mean spectral efficiency (SE) of the mmWave link between a random UE and its nearest AP -- the mean capacity of this link per $1$\,Hz.

{
\color{black}
We account for the LoS blockage probability, $p_{\text{B}}(x)$, by combining the model for the blockage by other pedestrians from~\cite{margo_tvt_mc} and the cone self-blockage model from~\cite{heath_self_blockage}, i.e.,
\begin{align}
p_{\text{B}}(x)=1-\frac{\left[2\pi-\arcsin{\left(\frac{r_\text{B}}{r_\text{B} + r_\text{U}}\right)}\right]}{2\pi} e^{-2r_{\text{B}}\lambda\big(\frac{x (h_{\text{B}}-h_{\text{U}})}{h_{\text{T}}}+r_{\text{B}}\big)},
\end{align}
where $x$ is the 2D distance between the UE and its nearest AP, while $h_{\text{T}}$ is the relative drone height compared to the UEs.
}

Observe that for a certain 2D distance of $x$ the SE, $S(x)$, is a mixture of two values that correspond to blocked and non-blocked link conditions (as detailed in subsection~\ref{sec:propagation}) weighed with probabilities, $p_{\text{B}}(x)$ and $1-p_{\text{B}}(x)$, i.e.,
\begin{align}\label{eqn:c_0}
\begin{split}
S(x) &= p_{\text{B}}(x) \log_{2}\left(1 + P_{\text{B}}(x) / (N_{0}N_{\text{F}})\right) +\\
&\left[1 - p_{\text{B}}(x)\right] \log_{2}\left(1 + P_{\text{N}}(x) / (N_{0}N_{\text{F}})\right),
\end{split}
\end{align}
where $N_{\text{F}}$ is the noise figure and $N_0$ is the thermal noise power.

Using (\ref{eqn:c_0}), we obtain the mean SE of the link between a randomly selected UE and its nearest UAV-based mmWave AP under airborne \textcolor{black}{($\overline{S}_{\text{A}}$) and landed ($\overline{S}_{\text{L}}$) deployments:
\begin{align}\label{eqn:mean_c_0}
\overline{S}_{\text{A}} = \int^{R}_{0} S(x) f_{X_{\text{A}}}(x) dx,\,\,\,\,\,\overline{S}_{\text{L}} = \int^{R}_{0} S(x) f_{X_{\text{L}}}(x) dx.
\end{align}}

\subsubsection{Network capacity}
Recall that there are $N_{\text{A}}$ and $N_{\text{L}}$ UAV-based mmWave APs continuously serving the UEs in airborne and landed cases, respectively. Hence, the mean network capacity for these options can be derived from (\ref{eqn:mean_c_0}) as\textcolor{black}{
\begin{align}\label{eqn:mean_network_c}
\overline{C}_{\text{A}} = N_{\text{A}} B \overline{S}_{\text{A}},\quad{} \overline{C}_{\text{L}} = N_{\text{L}} B \overline{S}_{\text{L}},
\end{align}}
where $B$ is the bandwidth of the employed mmWave channel.

\subsubsection{User capacity}
We finally determine the mean user capacity \textcolor{black}{-- the mean UE link capacity where all the radio resources are equally shared among all the UEs in the service area}: $\overline{C}_{\text{U},\text{A}}$ and $\overline{C}_{\text{U},\text{L}}$, respectively.
We first parameterize $p_{\text{B}}(x)$ not by the average density of the UEs in the area, $\lambda$, but {\color{black}by the current density of the UEs, $\lambda_{C} = K/(\pi R^2)$, \textcolor{black}{where $K$ is the number of UEs in the service area}:
\begin{align}
\hspace{-3mm}p_{\text{B}}(x, K)\hspace{-1mm}=\hspace{-1mm}1-\frac{\left[2\pi-\arcsin{\left(\frac{r_\text{B}}{r_\text{B} + r_\text{U}}\right)}\right]}{2\pi} e^{-2r_{\text{B}}\frac{K}{\pi R^2}\big(\frac{x (h_{\text{B}}-h_{\text{U}})}{h_{\text{T}}}+r_{\text{B}}\big)}\hspace{-1mm}.\hspace{-2mm}
\end{align}}

The user capacity, $C_{\text{U}}(x, K)$, for $M$ serving UAVs is then
\begin{align}\label{eqn:c_U}
\begin{split}
C_{\text{U}}(x, K) &= B\frac{M}{K}\Big[ p_{\text{B}}(x, K) \log_{2}\left(1 + P_{\text{B}}(x) / (N_{0}N_{\text{F}})\right) +\\
&\left[1 - p_{\text{B}}(x, K)\right] \log_{2}\left(1 + P_{\text{N}}(x) / (N_{0}N_{\text{F}})\right)\Big],
\end{split}
\end{align}
\textcolor{black}{where $M = N_{\text{A}}$ for the airborne option and $M = N_{\text{L}}$ for landed.}

Recall that $x$ and $K$ are independent random variables (RVs), while $K$ is discrete and follows a Poisson distribution with the mean of $\lambda \pi R^{2}$. Hence, $C_{\text{U}}(x)$ is a weighed sum of $C_{\text{U}}(x, K)$:
\begin{align}\label{eqn:c_U_x}
C_{\text{U}}(x) = \sum^{\infty}_{K=1} C_{\text{U}}(x, K) \frac{(\lambda \pi R^{2})^{K} e^{-\lambda \pi R^{2}}}{K!}.
\end{align}

Finally, the mean user capacity values under airborne and landed options, $\overline{C}_{\text{U},\text{A}}$ and $\overline{C}_{\text{U},\text{L}}$, are obtained similarly to (\ref{eqn:mean_c_0}):
\begin{align}\label{eqn:mean_c_U}
\overline{C}_{\text{U},\text{A}} = \int^{R}_{0}\left(\sum^{\infty}_{K=1} C_{\text{U}}(x, K) \frac{(\lambda \pi R^{2})^{K} e^{-\lambda \pi R^{2}}}{K!}\right) f_{X_{\text{A}}}(x) dx,\\
\overline{C}_{\text{U},\text{L}} = \int^{R}_{0}\left(\sum^{\infty}_{K=1} C_{\text{U}}(x, K) \frac{(\lambda \pi R^{2})^{K} e^{-\lambda \pi R^{2}}}{K!}\right) f_{X_{\text{L}}}(x) dx.
\end{align}

\section{Numerical Results}
\label{sec:numerical_results}
In this section, we collect our numerical results illustrating the trade-offs between the use of airborne and landed drones. We first study the UAV operation cycle stages separately in Fig.~\ref{fig:intermediate} and then focus on the overall system performance in Fig.~\ref{fig:final}. The major parameters are summarized in Table~\ref{tab:notation}.

\subsection{Analysis of Serving and En-route Stages}

\subsubsection{The effect of the drone height}

We start with Fig.~\ref{fig:plot1}, which presents the mean SE as a function of the drone height during the serving stage for $M = 5$. For both deployment options, we notice a clear ``optimal'' drone height featured by the highest SE: too low height results in numerous blockage events, while the opposite leads to an SNR degradation at the mmWave link due to the increased UE-UAV distance.

We then observe that the ``optimal'' heights for airborne and landed cases are different; hence, these options should not be compared with identical drone heights. Instead, for all further figures, we set the drone height to its ``optimal'' value, chosen individually for landed and airborne options. We also see a match between the results derived with our framework and those produced via computer simulations of our system model.

{
\color{black}
\begin{figure*}[h!]
\centering
	\begin{subfigure}[c]{0.325\textwidth}
		\centering
		\includegraphics[height=1.95in]{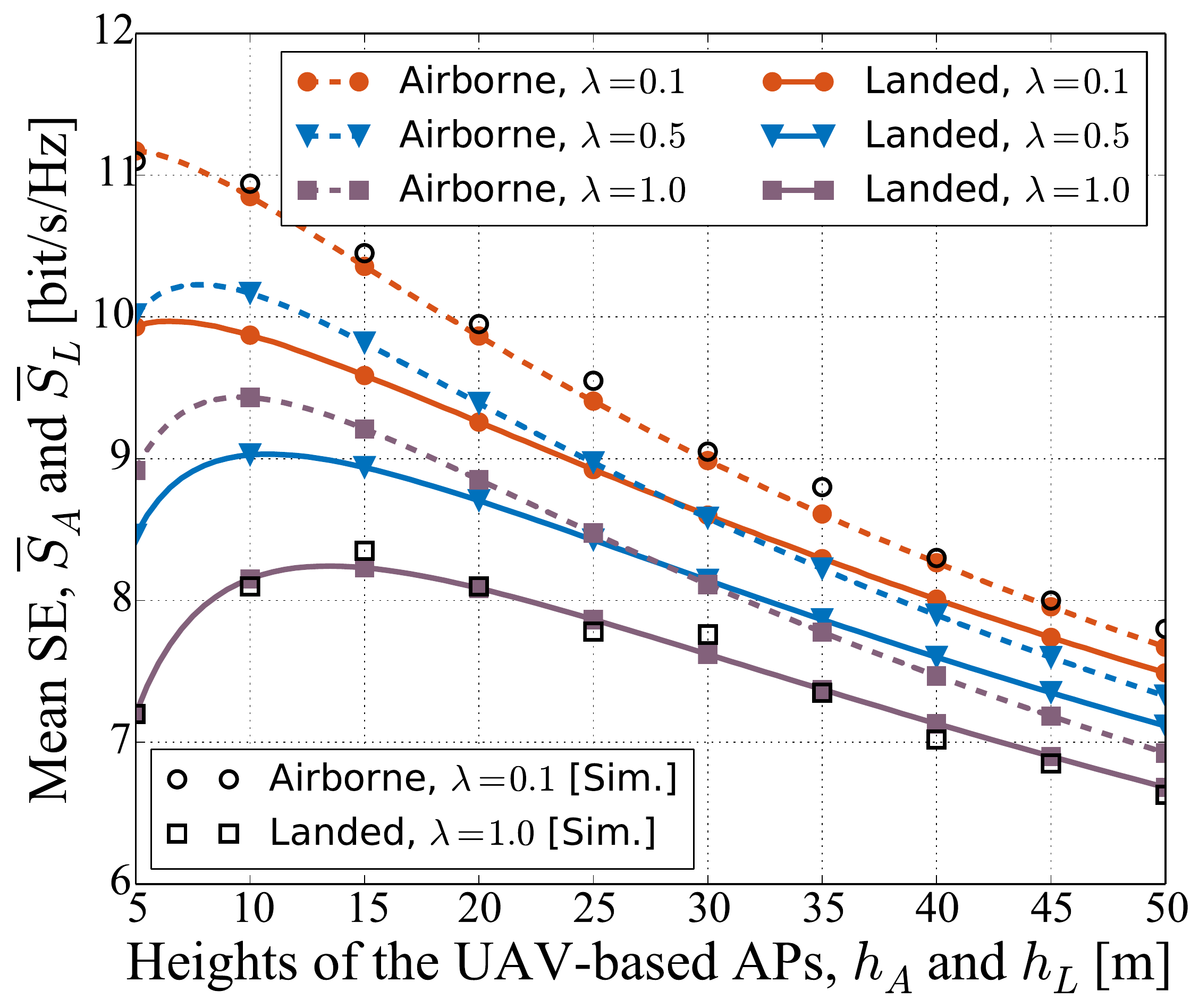}
	\caption{\textcolor{black}{Effect of the height of the UAV-based AP}}
	\label{fig:plot1}
	\end{subfigure}
	\begin{subfigure}[c]{0.325\textwidth}
		\centering
		\includegraphics[height=1.95in]{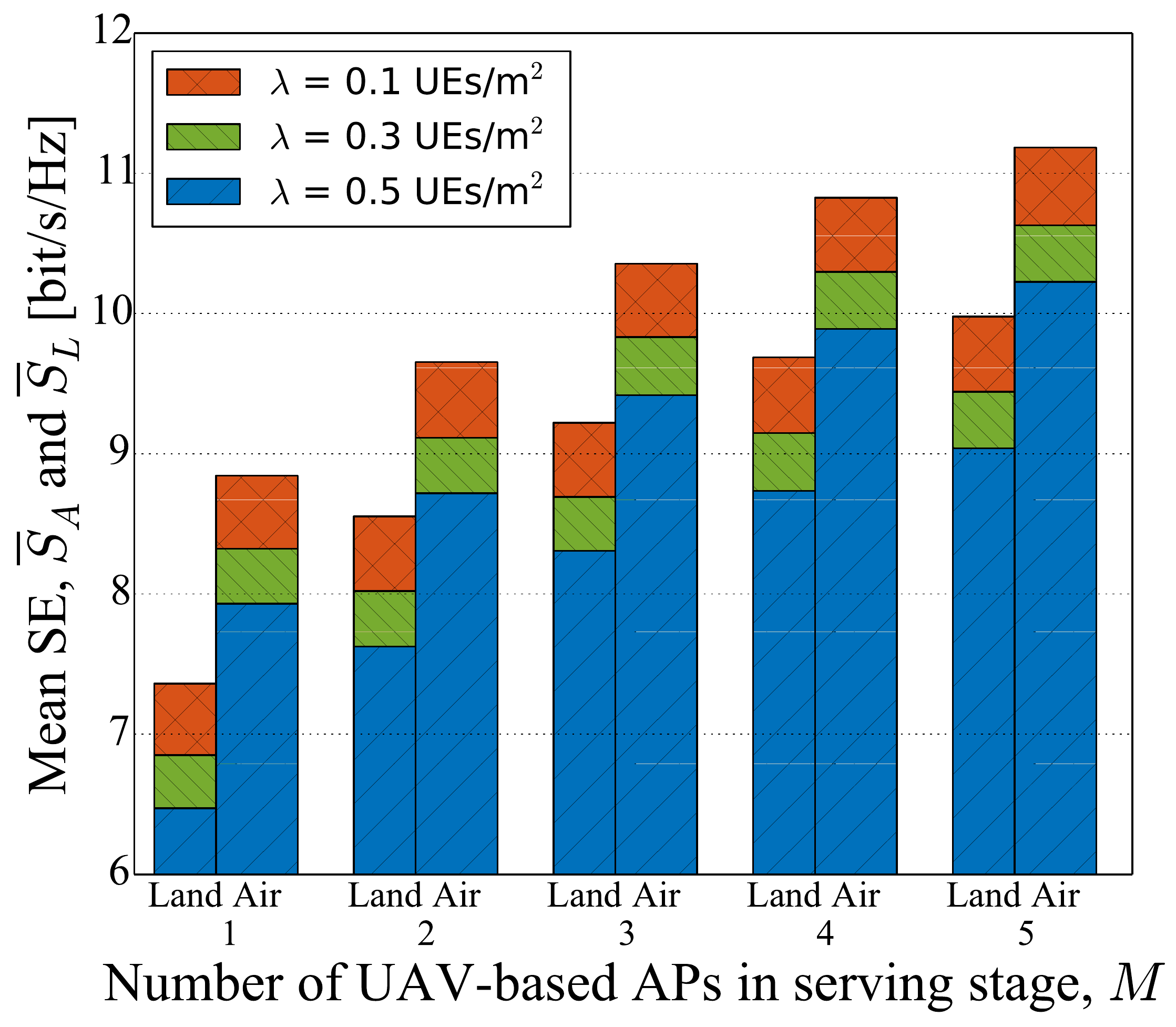}
	\caption{\textcolor{black}{Effect of the number of serving UAVs}}
	\label{fig:plot2}
	\end{subfigure}
	\begin{subfigure}[c]{0.325\textwidth}
		\centering
		\includegraphics[height=1.95in]{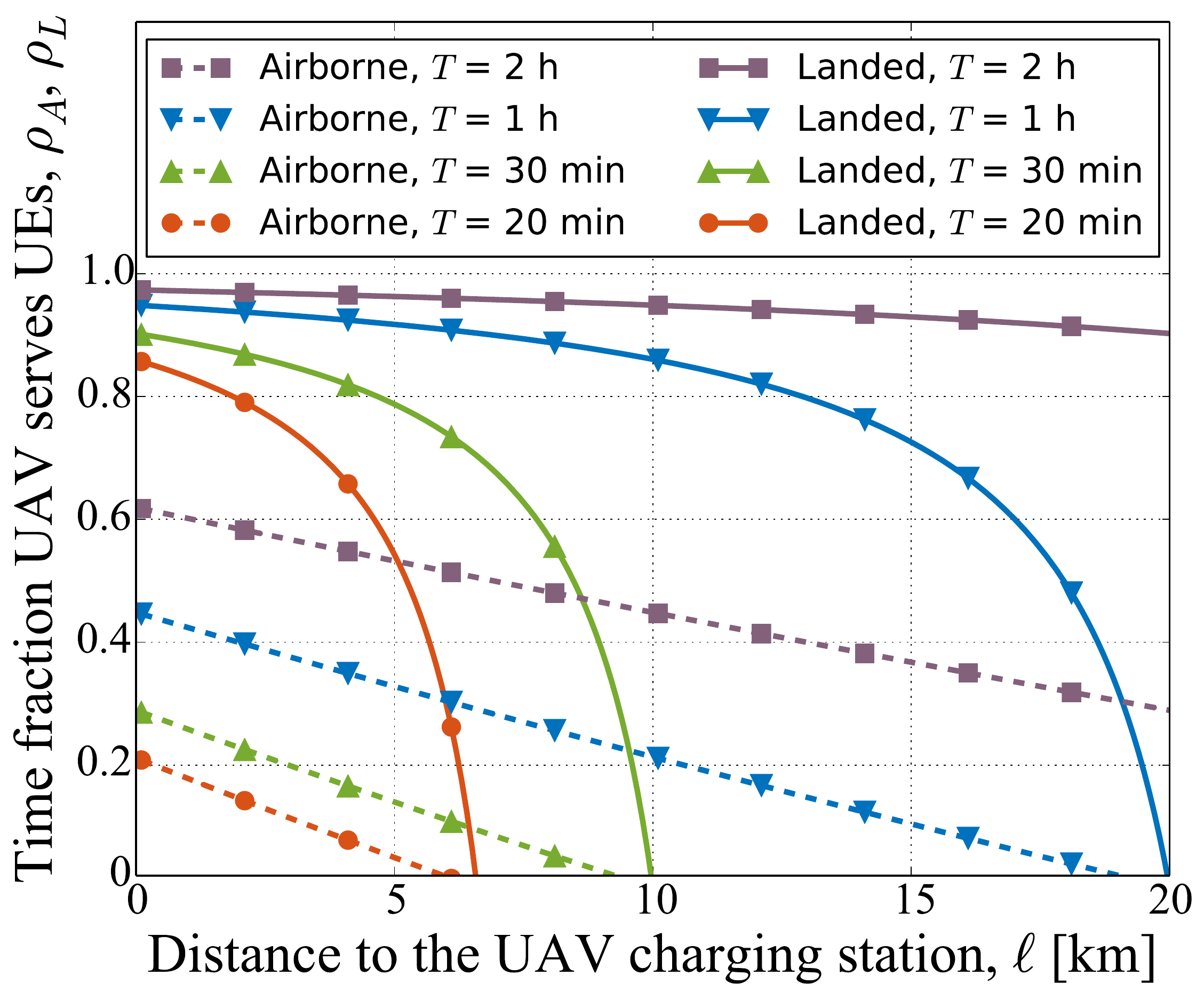}
	\caption{\textcolor{black}{Effect of the distance to the charging station}}
	\label{fig:plot3}
	\end{subfigure}
\caption{\textcolor{black}{Individual analysis of the \emph{serving} stage, Fig.~\ref{fig:plot1} and Fig.~\ref{fig:plot2}, and the \emph{en-route}$+$\emph{charging} stages, Fig.~\ref{fig:plot3}.}}
\label{fig:intermediate}
\end{figure*}

\begin{figure*}[h!]
\centering
	\begin{subfigure}[c]{0.325\textwidth}
		\centering
		\includegraphics[height=1.95in]{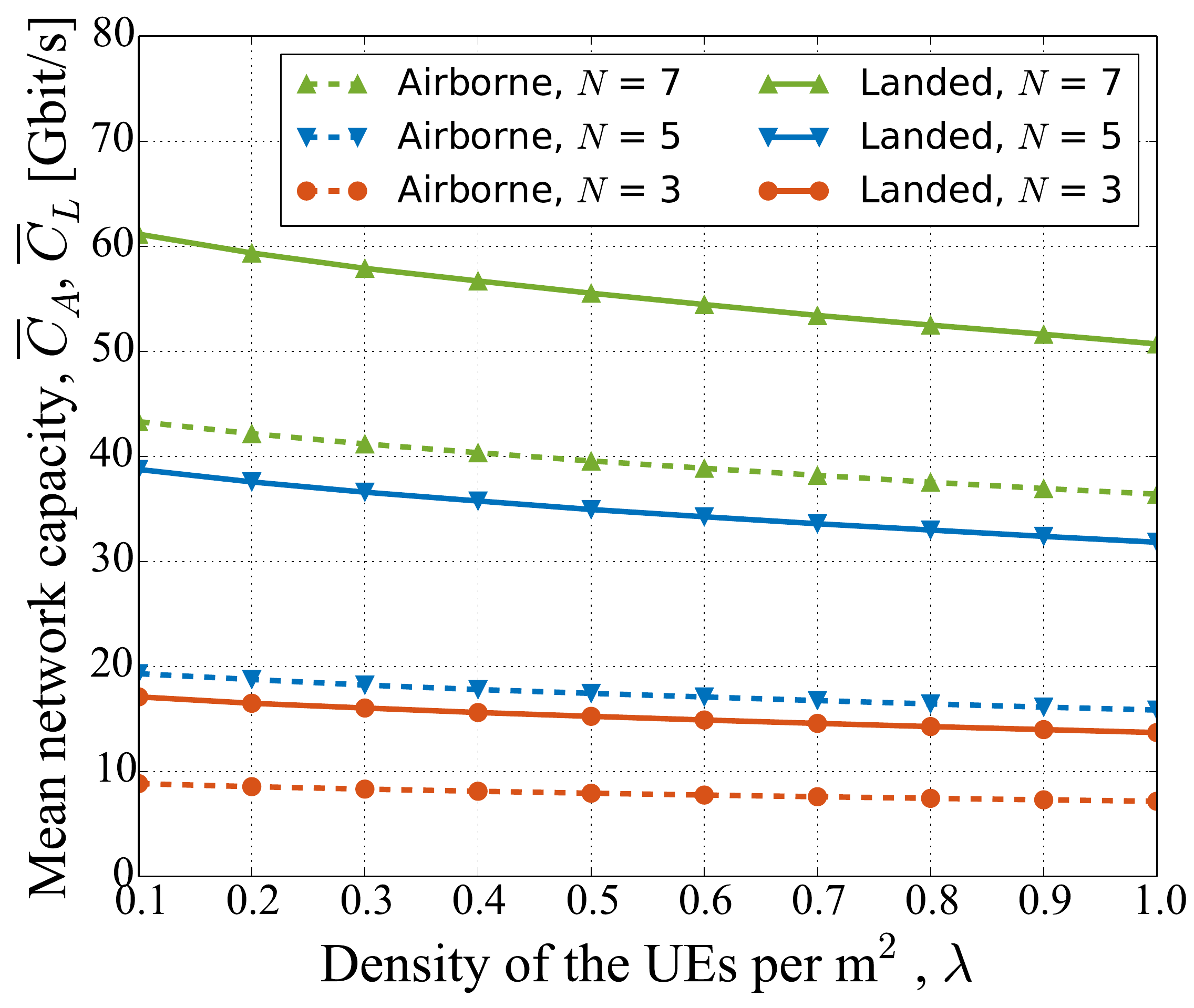}
	\caption{\textcolor{black}{Effect of the UE density}}
	\label{fig:plot4}
	\end{subfigure}
	\begin{subfigure}[c]{0.325\textwidth}
		\centering
		\includegraphics[height=1.95in]{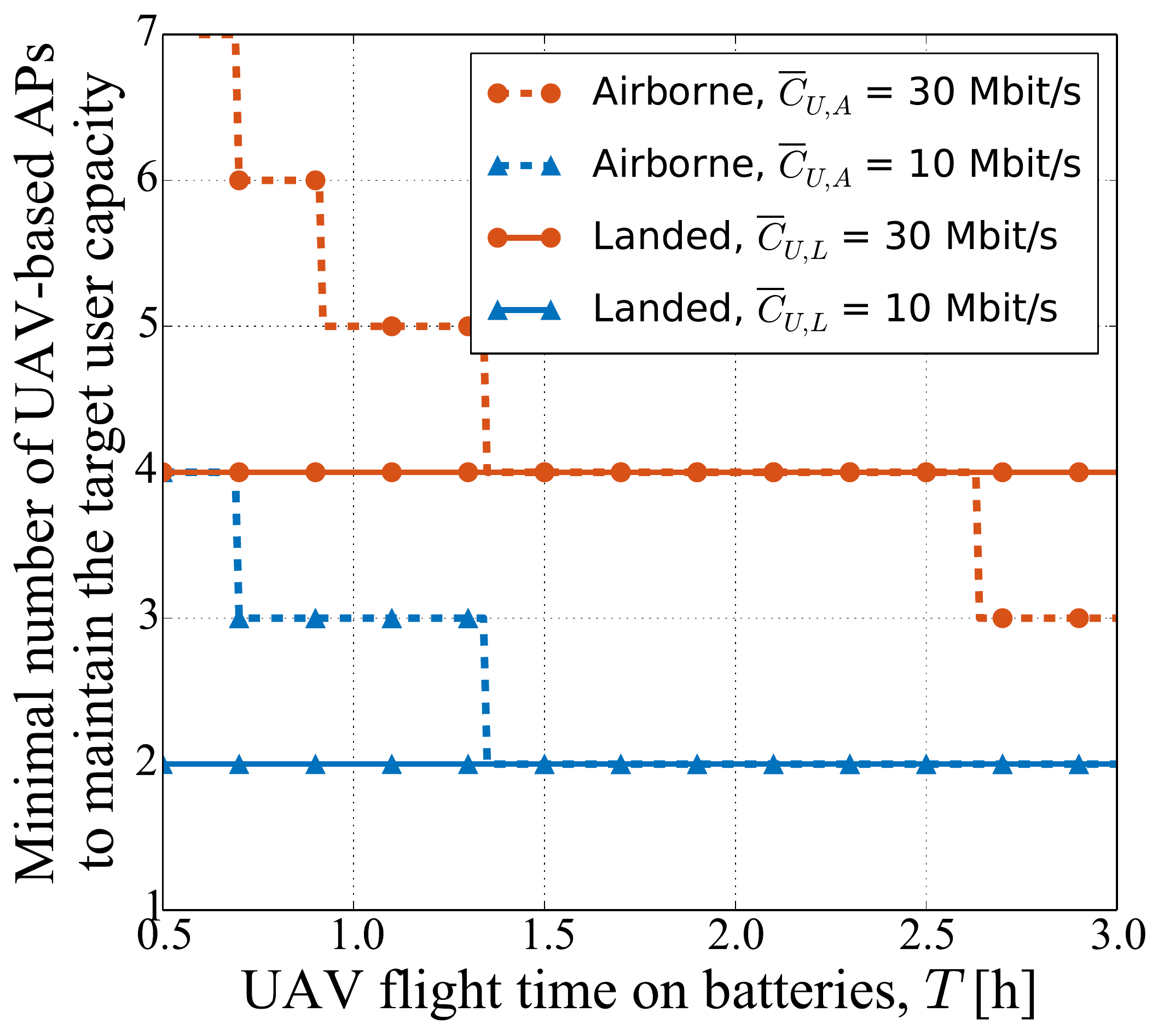}
	\caption{\textcolor{black}{Effect of the flight time on batteries}}
	\label{fig:plot5}
	\end{subfigure}
	\begin{subfigure}[c]{0.325\textwidth}
		\centering
		\includegraphics[height=1.95in]{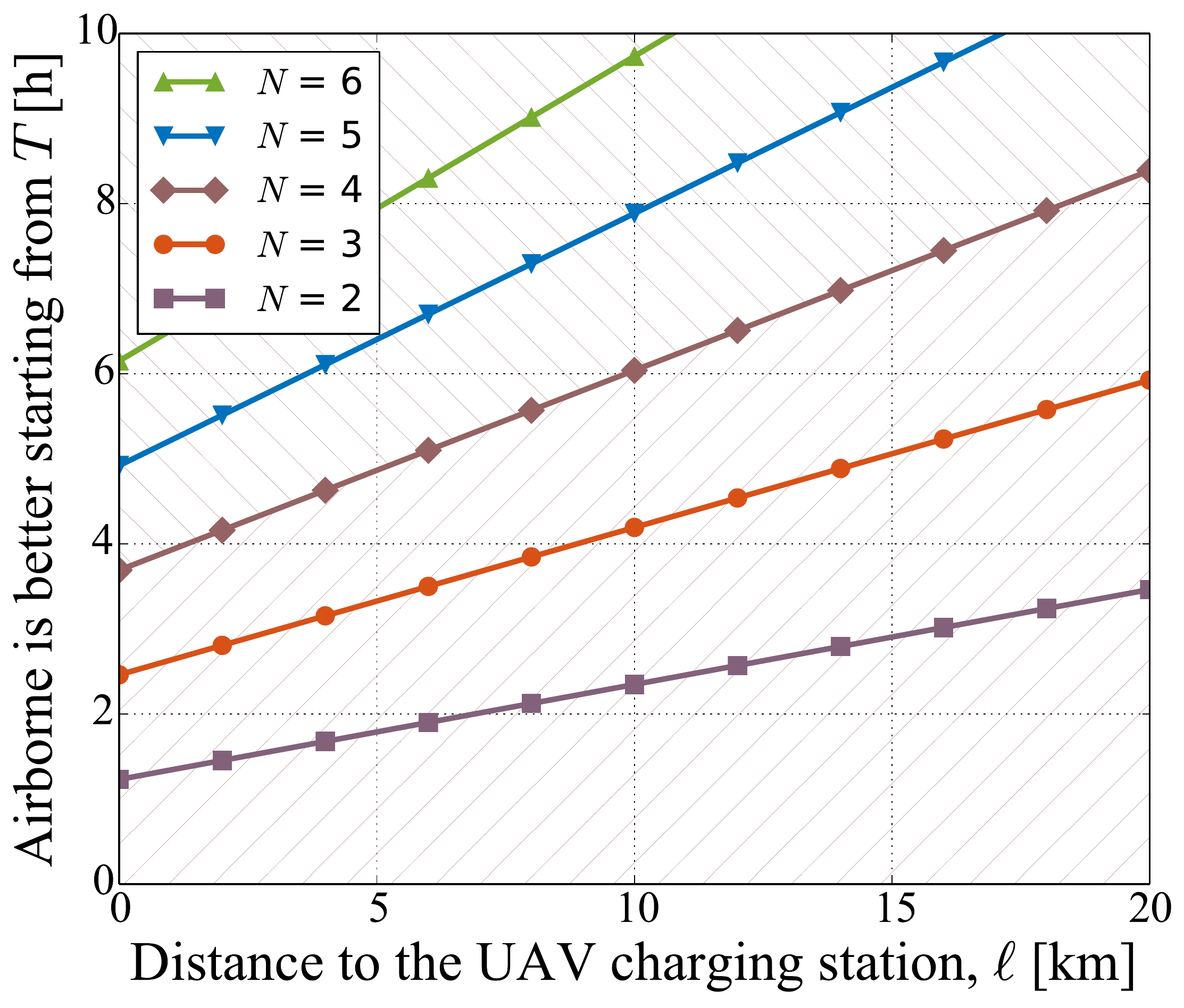}
	\caption{\textcolor{black}{Key trade-offs for ``Airborne vs. Landed''}}
	\label{fig:plot6}
	\end{subfigure}
\caption{\textcolor{black}{Comparing the performance of \emph{airborne} vs. \emph{landed} deployment options at the system level.}}
\label{fig:final}
\end{figure*}
}

\subsubsection{The effect of the number of serving drones}

We proceed with Fig.~\ref{fig:plot2}, which displays the mean SE as a function of the number of drones in the \emph{serving stage}, $M$. Here, we clearly observe that the airborne option results in up to \textcolor{black}{$20\%$} higher SE than the landed case across all the numbers of serving drones available and all the densities of the UEs in the area. 

\subsubsection{The effect of the distance to the charging station}

The opposite trend is observed in Fig.~\ref{fig:plot3}, which illustrates the \emph{en-route+charging stages}, where the shares of time when the selected drone can serve the UEs -- $\rho_{\text{A}}$ and $\rho_{\text{L}}$ for airborne and landed cases, respectively -- are functions of the distance between the service area and the drone charging station, $\ell$. \textcolor{black}{Since the landed drones are featured by a considerably reduced power consumption in the serving stage (the engines are off), $\rho_{\text{L}}$ is $50\%$--$400\%$ greater than $\rho_{\text{A}}$ for the same parameters.}

\begin{table}[!t]
 \caption{\\Notation and parameters of our numerical study}
 \label{tab:notation}
 \begin{center}
 {\scriptsize
     \begin{tabular}{p{0.09\columnwidth}p{0.53\columnwidth}p{0.21\columnwidth}}
    \hline
    \textbf{Symbol}&\textbf{Definition}&\textbf{Value}\\
    \hline
    \multicolumn{2}{l}{\quad{}\quad{}\textbf{mmWave AP service parameters}}\\
    \hline
    $f_{\text{C}}$,\,\,\,$B$ & Carrier frequency and system bandwidth & $28$\,GHz, $1$\,GHz\\
    $P_{\text{A}}$ & UAV-based mmWave AP transmit power & $23$\,dBm\\
    $A_{N}$/$A_{B}$ & Power reduction due to human body blockage & $20$\,dB~\cite{haneda_blockage_vtc}\\
    $G_{\text{A}}$, $G_{\text{U}}$ & AP and UE antenna gains & $15$\,dB, $5$\,dB\\
    $N_{0}$, $N_{\text{F}}$ & Noise power over $1$\,GHz band and noise figure & $-84$\,dBm, $5$\,dB\\
    \hline
    \multicolumn{2}{l}{\quad{}\quad{}\textbf{UAV flight parameters}}\\
    \hline
    $\nu$,\,\,\,$T_{\text{C}}$& UAV cruise flight speed and full charging time & $40$\,km/h, $1$\,h\\
    $P_{\text{E}}$,\,\,\,$P_{\text{H}}$& UAV engines power during flight and hovering & $871$\,W, $1024$\,W~\cite{revision_energy_all}\\
    $P_{\text{T}}$& Total power of UAV-based mmWave AP & $47$\,W~\cite{revision_uav_ap_power}\\
    $R$ & Radius of the considered service area & $50$\,m\\
    \hline
    \end{tabular}
 }
 \end{center}
 \end{table}

\subsection{System-Level Analysis}

\subsubsection{The effect of the human crowd density}
We now proceed with Fig.~\ref{fig:final}, which studies the full cycle of drone operation. Particularly, Fig.~\ref{fig:plot4} presents the mean network capacity as a function of the UEs density. \textcolor{black}{Here, the landed drones enable $10\%$--$40\%$ higher capacity than the airborne case across all the UE densities, $\lambda$, and the total numbers of drones, $N$.} Hence, the advantage of the landed option during the en-route stage illustrated in Fig.~\ref{fig:plot3} is more pronounced than the benefit of the airborne option during the serving stage, as in Fig.~\ref{fig:plot2}.

\subsubsection{The effect of the flight time on batteries}
It is important to note that the conclusions from Fig.~\ref{fig:plot4} are straightforward only when the UAVs are equipped with relatively small batteries, which allow up to one hour of flight time ($T = 1$\,h). We study the effect of battery lifetime in more detail in Fig.~\ref{fig:plot5}, by presenting the minimum number of drones, $N$, required to maintain the desired UE capacity level as a function of $T$.

As can be observed in Fig.~\ref{fig:plot5}, the landed option remains preferable for most of the considered setups, when $T \leq 1.3$\,h. Further, the same number of drones is required for both options when \textcolor{black}{$T \in [1.3$\,h, $2.7$\,h$)$. Finally, for $\overline{C}_{\text{U},\text{A}} = \overline{C}_{\text{U},\text{L}} = 30$\,Mbit/s and $T \geq 2.7$\,h, the airborne case becomes preferable: $3$ airborne UAVs required vs. $4$ drones with the landed option.}

\subsection{Summary: Airborne vs. Landed Cases}

We conclude with Fig.~\ref{fig:plot6}, which highlights the regions for the ``airborne vs. landed'' trade-offs, when the mean network capacity is the characteristic of interest. \textcolor{black}{Recall that the exact mean network capacity depends on multiple factors: the drone height, the density of the UEs in the area, etc. In contrast, as in Fig.~\ref{fig:plot6}, the general answer to ``Which option results in higher network capacity?'' is determined primarily by: (i) the distance to the charging station, $\ell$; (ii) the UAV flight time on batteries, $T$; and (iii) the total number of drones, $N$.}

\textcolor{black}{Particularly, if four UAVs are available ($N=4$), all the system setups laying above the corresponding curve in Fig.~\ref{fig:plot6} are better served with the airborne deployment, while all the sets below this curve achieve higher capacity with the landed option.} In more detail, we notice that the landed case is more attractive in complex setups: high $N$, high $\ell$, and low $T$. 

\section{Conclusions}
\label{sec:conclusions}
In this letter, we developed a mathematical framework to compare the capacity of the drone-cells deployed on-demand by \emph{airborne} and \emph{landed} UAV-based mmWave APs. This framework was then applied to identify the setups where one of the deployment choices is preferable over the other. Our analysis particularly indicates that, while the mean capacity is subject to a number of factors, the preferred option (\emph{airborne} vs. \emph{landed}) mainly depends on \textcolor{black}{three parameters}: (i)~the separation distance between the service area and the drone charging station; (ii)~the UAV battery lifetime, and (iii)~the number of UAV-based mmWave APs in use. The proposed approach can be further extended to account for a presence of terrestrial mmWave APs, \textcolor{black}{multiple drone charging locations}, as well as for different considerations regarding the UAV backhaul.

\balance
\bibliographystyle{IEEEtran}
\bibliography{static_drones_short_blue}

\begin{thebibliography}{10}
\providecommand{\url}[1]{#1}
\csname url@samestyle\endcsname
\providecommand{\newblock}{\relax}
\providecommand{\bibinfo}[2]{#2}
\providecommand{\BIBentrySTDinterwordspacing}{\spaceskip=0pt\relax}
\providecommand{\BIBentryALTinterwordstretchfactor}{4}
\providecommand{\BIBentryALTinterwordspacing}{\spaceskip=\fontdimen2\font plus
\BIBentryALTinterwordstretchfactor\fontdimen3\font minus
  \fontdimen4\font\relax}
\providecommand{\BIBforeignlanguage}[2]{{%
\expandafter\ifx\csname l@#1\endcsname\relax
\typeout{** WARNING: IEEEtran.bst: No hyphenation pattern has been}%
\typeout{** loaded for the language `#1'. Using the pattern for}%
\typeout{** the default language instead.}%
\else
\language=\csname l@#1\endcsname
\fi
#2}}
\providecommand{\BIBdecl}{\relax}
\BIBdecl

\bibitem{mmWave_survey}
X.~{Wang}, L.~{Kong}, F.~{Kong}, F.~{Qiu}, M.~{Xia}, S.~{Arnon}, and G.~{Chen},
  ``Millimeter wave communication: {A} comprehensive survey,'' \emph{IEEE
  Commun. Surveys \& Tut.}, vol.~20, no.~3, pp. 1616--1653, Q3 2018.

\bibitem{mmWave_uav_magazine2}
P.~{Yu} \emph{et~al.}, ``Capacity enhancement for {5G} networks using mm{W}ave
  aerial base stations: {S}elf-organizing architecture and approach,''
  \emph{IEEE Wireless Commun.}, vol.~25, no.~4, pp. 58--64, Aug. 2018.

\bibitem{related_uav_coverage}
A.~{Trotta} \emph{et~al.}, ``Joint coverage, connectivity, and charging
  strategies for distributed {UAV} networks,'' \emph{IEEE Trans. on Robotics},
  vol.~34, no.~4, pp. 883--900, Aug. 2018.

\bibitem{perch_uav_1}
K.~Hang \emph{et~al.}, ``Perching and resting -- {A} paradigm for {UAV}
  maneuvering with modularized landing gears,'' \emph{Science Robotics},
  vol.~4, no.~28, Mar. 2019.

\bibitem{margo_tvt_mc}
M.~{Gapeyenko} \emph{et~al.}, ``{On the Degree of Multi-Connectivity in {5G}
  Millimeter-Wave Cellular Urban Deployments},'' \emph{{IEEE Transactions on
  Vehicular Technology}}, vol.~{68}, no.~{2}, pp. {1973--1978}, {Feb.} {2019}.

\bibitem{related_heath_conference}
R.~{Shinkuma} and Y.~{Goto}, ``Wireless multihop networks formed by unmanned
  aerial vehicles with separable access points and replaceable batteries,'' in
  \emph{Proc. of IEEE UEMCON}, Oct. 2016, pp. 1--6.

\bibitem{related_spatial_framework}
W.~{Yi} \emph{et~al.}, ``A unified spatial framework for {UAV}-aided mm{W}ave
  networks,'' \emph{IEEE Trans. on Commun.}, vol.~67, no.~12, pp. 8801--8817,
  Dec. 2019.

\bibitem{revision_energy_all}
H.~{Shakhatreh} \emph{et~al.}, ``{On the continuous coverage problem for a
  swarm of {UAV}s},'' in \emph{{Proc. of IEEE Sarnoff Symposium}}, {Sep.}
  {2016}, pp. {130--135}.

\bibitem{3gpp_los}
{3GPP}, ``Study on channel model for frequencies from 0.5 to 100 {GH}z
  ({R}elease 16),'' {3GPP TR 38.901 V16.0.0}, Oct. 2019.

\bibitem{haneda_blockage_vtc}
K.~Haneda \emph{et~al.}, ``5{G} {3GPP}-like channel models for outdoor urban
  microcellular and macrocellular environments,'' in \emph{Proc. of IEEE VTC
  Spring}, May 2016, pp. 1--7.

\bibitem{davies1979cluster}
D.~L. Davies and D.~W. Bouldin, ``A cluster separation measure,'' \emph{IEEE
  Trans. on Pat. Analys. and Mach. Intel.}, no.~2, pp. 224--227, Apr. 1979.

\bibitem{circle_packing_saad}
M.~{Mozaffari}, W.~{Saad}, M.~{Bennis}, and M.~{Debbah}, ``Efficient deployment
  of multiple unmanned aerial vehicles for optimal wireless coverage,''
  \emph{IEEE Comm. Let.}, vol.~20, no.~8, pp. 1647--1650, Aug. 2016.

\bibitem{revision_link_projection}
A.~M. Mathai, \emph{{An introduction to geometrical probability:
  {D}istributional aspects with applications}}.\hskip 1em plus 0.5em minus
  0.4em\relax {CRC Press}, {1999}.

\bibitem{parker1996deterministic}
R.~G. Parker, \emph{Deterministic scheduling theory}.\hskip 1em plus 0.5em
  minus 0.4em\relax CRC Press, 1996.

\bibitem{heath_self_blockage}
T.~{Bai} and R.~W. {Heath}, ``{Analysis of self-body blocking effects in
  millimeter wave cellular networks},'' in \emph{{Proc. of ASILOMAR}}, {Nov.}
  {2014}, pp. {1921--1925}.

\bibitem{revision_uav_ap_power}
M.~A. Imran and E.~Katranaras, ``{{Energy efficiency analysis of the reference
  systems, areas of improvements and target breakdown}},'' {INFSO-ICT-247733
  EARTH}, Tech. Rep., {Jan.} {2012}.

\end{thebibliography}

\end{document}